\documentclass{llncs}

\usepackage{adjustbox}
\usepackage{caption}
\usepackage[hyphens]{url}
\usepackage{hyperref}
\usepackage{array}

\providecommand{\keywords}[1]
{
  \small	
  \textbf{\textit{Keywords--}} #1
}

\usepackage{float}
\begin{document}
\title{\Large \bf Text Generation for Dataset Augmentation in Security
Classification Tasks}

\author{%
{\rm Anonymous Authors}\\
Anonymous Institution
}

\author{
{\rm Alexander P.\ Welsh} \and {\rm Matthew Edwards}\\
School of Computer Science\\
University of Bristol\\
Bristol, UK\\
} 

\maketitle

\begin{abstract}  

Security classifiers, designed to detect malicious content in computer systems
and communications, can underperform when provided with insufficient training
data.  In the security domain, it is often easy to find samples of the negative
(benign) class, and challenging to find enough samples of the positive
(malicious) class to train an effective classifier.  This study evaluates the
application of natural language text generators to fill this data gap in
multiple security-related text classification tasks. We describe a variety of
previously-unexamined language-model fine-tuning approaches for this purpose and
consider in particular the impact of disproportionate class-imbalances in the
training set.



Across our evaluation using three state-of-the-art classifiers designed for
offensive language detection, review fraud detection, and SMS spam detection, we
find that models trained with GPT-3 data augmentation strategies outperform both
models trained without augmentation and models trained using basic data
augmentation strategies already in common usage. In particular, we find
substantial benefits for GPT-3 data augmentation strategies in situations with
severe limitations on known positive-class samples.

\end{abstract}



\section{Introduction}

Detecting malicious activity has been a task tackled by machine learning 
classifiers since the 1980s~\cite{MLFraudDetectionOrigins}. Classifiers are
first trained with datasets of both positive (malicious) and negative (benign)
samples, and then evaluated in settings that should reflect their intended
deployment scenario~\cite{arp2022and}.  It is well known that classifiers will
perform better with larger datasets, as long as the data is not of lesser
quality~\cite{moredata1,moredata2}. This can sometimes be a problem when
working in the security domain; malicious content like fraud, malware delivery,
and offensive comments are generally a small minority of all data, and cases are
often
under-reported~\cite{underreporting2,chequeFraudUnderreporting,twitterbans}.
This means that while negative cases are often plentiful, a lack of positive
cases can be a limitation.

Dataset augmentation is a technique used to artificially expand the size of a
dataset by creating new samples based on the data available~\cite{augmentation}.
Existing techniques for augmenting text data include swapping words for synonyms
or translating a sentence to another language and back
again~\cite{genAug2kumar}. These techniques are intended to produce small
differences without changing the fundamental nature of the data. These methods
have substantial limitations; lacking a complex understanding of the meaning and
structure of the data they are augmenting, they often either alter the original
samples very little or produce heavily-distorting mutations that may prevent
them from being considered a true example of the intended class.

This paper posits that security classifiers can be improved by creating new
positive training samples with a text-generating large language model, and
especially in cases where true positive data is limited. Modern language models
are good at writing text that is logical and coherent. They can often be tuned
with a small number of samples on a given topic. This allows the generator to
match the desired style, and focus on a certain subject. In a situation where data is
limited, the few true samples that are available could be used to automatically
create more in a similar style.

Our aim for this technique is that the generated samples will be of high enough
quality to measurably improve the performance---on unseen data---of security
classifiers that use them as training data.  
This, in turn, would help produce
more accurate detectors for malicious content. In particular, this technique
could greatly support security classification tasks in situations where there is
restricted access to samples, such as small enterprises attempting to build
security into their systems or software, or law enforcement agencies
investigating rare or underreported crime. 
 While other work has discussed this
technique more generally in natural language processing (NLP)
tasks~\cite{genAug1,genAug2kumar,genAug3}, prior research has not considered
elements key to the context of security-focused tasks.  Data-limited scenarios
have been examined, but not imbalanced and adversarial datasets, and the
fine-tuning process required has not been rigorously explored. There has also
been significant progress in text generation techniques since prior
investigations took place, presenting as-yet unevaluated opportunities.

To assess the viability of our approach, we apply a large language model to
augment datasets for three different detection tasks.  
In each case we first
replicate current best-performing classifiers from the literature. We then
present experiments demonstrating the impact of limiting the availability of
positive-class data, and show how augmenting the training dataset with a modern
large language model can repair this impact.

Our overall contributions include:
\begin{enumerate} \item We evaluate modern text generation as a method of
dataset augmentation for malicious content classification tasks. We find that,
with a minor exception, models trained with such data augmentation outperform
both models lacking augmentation and models using more widely-used basic data
augmentation strategies.

\item We provide what we believe is the first investigation into fine-tuning
text generators in the context of \emph{disproportionately}
data-limited datasets, including the effect of different levels of data
reduction. We find that modern text generators can be especially effective for
data augmentation in cases where positive-class samples are least
available.

\item We assess different methods of fine-tuning language models for text
generation as a method of dataset augmentation. We find mixed results pointing
to possible tradeoffs between strategies depending on classifier selection and
the quantity of data available for fine-tuning.

\end{enumerate}

The remainder of this paper proceeds as follows. In Section~\ref{sec:background}
we provide a brief introduction to text generation and natural language classification within
security, and highlight closely related work within the field. 
In Section~\ref{sec:execution} we outline our experimental process for
data limitation, fine-tuning and data augmentation. Section~\ref{sec:results}
presents the results of applying this process to three security
classification tasks. We discuss our observations and the implications and
limitations of our method in Section~\ref{sec:analysis}, before concluding
with our key takeaways for future work.

\section{Background}
\label{sec:background}

\subsection{Dataset Augmentation in NLP}\label{sectionDA}
A lack of data presents a major problem for security classification tasks.
Building effective classifiers often requires large amounts of data which can only be collected after an incident has taken place. This means that before these filtration systems become fully functional, the public is repeatedly put at risk. Researchers have long been looking at how to lower the number of real-world samples needed before it is possible to identify patterns. 
It is widely known that classifiers with a learning component will perform better with larger datasets, given the data is not of lesser quality~\cite{moredata1,moredata2}. This is because it allows a classifier to better generalise to a given class. Limited data can cause overfitting.

Dataset augmentation techniques are strategies to artificially expand the size
of a dataset, by creating new samples based on the ones
available~\cite{augmentation}. For example, a data scientist using a set of
pictures may rotate, crop, or hue shift images by a small offset and add these
new samples to their set. These samples are just edited versions of existing
data, but can provide additional information to the model. When done correctly, these techniques help to reduce overfitting and improve performance~\cite{augmentationOverfitting} by providing more generalised data.

Dataset augmentation for textual data is arguably more challenging due to the
complex grammar of human languages. Small edits to punctuation, word choice, or
structure can completely change the meaning of a sentence. Even so, simple
techniques such as the random removal or reordering of words can measurably
improve performance~\cite{textDataAugPerformance}.

\subsubsection{Synonym Replacement}
Synonym replacement is a computationally complex strategy but has a simple concept. Words are extracted at random from a sample and replaced with either synonyms or closely related words. For example, \emph{cat} could become the synonym \emph{feline} or the closely related word \emph{dog}. In many cases, the result is a new sample with the same meaning but slightly different word choices. This however is not a perfect strategy as some swaps can change a sample entirely. A \emph{two-dimensional \textbf{plane}} could be very different to a \emph{two-dimensional \textbf{aircraft}}.

\subsubsection{Word Insertion} \label{sectionwi}
Word insertion is a technique that can be performed in several ways. With a random approach, words from a dictionary could be placed at any point in a sample. It is however common to use more advanced techniques. These fall into two major categories. One is to use word embeddings, the other is to use a language model.
Word embeddings are representations of words, usually taking the form of a vector. These are assigned in such a way that similar words are closer to each other in the vector space are more related to each other. These can be used to intelligently insert more relevant words at points throughout a sample.
Language models can also be used for this purpose. Given a sample, `blank' words can be inserted at random points, and the language model can be used to predict what the words should be. Given that language models are often trained by blanking out words from real sentences, they can perform well at this task.

\subsection{Text Generation for Dataset Augmentation}
The first use of text generation as a form of dataset augmentation was by Wu et al.~\cite{genAugBERT}. They introduced CBERT, a language model based on BERT~\cite{BERT} designed specifically for dataset augmentation, which worked by deterministically blanking out words in true samples and filling in the gaps with the language model.

Fully generating samples as augmentation data was first presented by Anaby-Tavor et al.~\cite{genAugFirst}.
Their methodology is to fine-tune a pre-trained GPT-2 model with a base dataset. This is then used to generate a set of samples 10$\times$ greater than the original dataset.
They take advantage of the fact that the classifier they are using reports confidence levels as well as a class label. 
Any sample with a confidence level below a certain threshold is dropped, and the remainder are added to the training set. They were able to show improvements of 2\% to 58\% depending on the classifier and classification task. A criticism of this task is that while they compared the performance of their generator-augmented datasets to basic techniques, they did not extend the filtration step to other methods. This may have unfairly biased their results.

Kumar et al.~\cite{genAug2kumar} present a more general method. They do not include a filtration step, instead directly using the language model output.
They also extend testing to three language models, each constructed in a different style.
They opted to work on three different tasks, namely sentiment classification~\cite{kmrData1}, intent classification~\cite{kmrData2}, and question topic classification~\cite{kmrData3}. GPT-2, the largest model tested, did not generally perform as well as the other two models. It was able to produce ``very coherent" text, however the class was not translated well into the samples it generated. 
One notable result of this investigation was unusually high effectiveness when working in low-data scenarios.

Quteineh et al.~\cite{genAug1} further investigate the use of generator augmentation while working with extremely small datasets of only 5 or 10 samples per class. They present methods for highly effective use of this data, emphasising efficiency. These results are again verified on varying tasks, and the authors note that the methods described should apply to any domain or language, given a suitable language model. It is worth noting that they do however make use of manual labelling. 

With the introduction of GPT-3~\cite{GPT-3} in 2020, researchers were given
access to a much more powerful model. Yoo et al.~\cite{GPT-3-Mix} showed
impressive improvements with this new class of generator. They use a different
labelling methodology with continuous `soft' class labels, rather than discreet
`hard' classes. This helps with the transfer of knowledge throughout the models
by providing effectively what is a measure of confidence with each sample. This
made their implementation more in line with the style of Anaby-Tavor et al.~\cite{genAugFirst}, rather than Kumar et al.~\cite{genAug2kumar}.

Initially, we believed our research may be the first to apply a
modern language model, such as GPT-3, for dataset augmentation with discrete
class labels. This topic has been recently investigated by Sahu et
al.~\cite{GPT-3-aug-recent}. However, they focus on different aspects of the
problem, such as sampling parameters, and apply the the technique in the field
of intent classification, with generally high numbers of classes (7, 64, 77, and
150 for each of their 4 chosen datasets respectively).

We investigate how this idea can be applied to the specific domain of
data-limited security classification. This area deals with similar language to
many standard NLP classification tasks. Security classification can often be
reduced to other fields such as a combination of sentiment and intent
classification. These tasks have been well researched in recent
years~\cite{snipsSOTA,snipsSOTA2,snipsSOTA3,SST-2SOTA,SSt-2SOTA2,SSt-2SOTA3}.
However, data-limitation in particular has only been explored in a proportionate
setting, with a lack of research exploring heavily imbalanced scenarios where one class 
is common and others are rare -- a scenario that is commonly the case in the 
classification of malicious behaviour.


\section{Method}
\label{sec:execution}

\subsection{Task \& Dataset Selection}

We selected three different text-based security and online harm tasks to
evaluate our dataset augmentation approach. In each case, we first replicated
recent classifiers from the literature before measuring the effect of data loss
and the viability of augmentation solutions. The implementations we selected
were: offensive language detection by Dai et al.~\cite{olid-task-paper},
deceptive review detection by Salunkhe~\cite{reviews-task-paper}, and SMS spam
detection, by Chandra \& Khatri~\cite{sms-task-paper}.

\subsubsection{Task 1: Offensive Language Detection} 

Given the scale of modern social media, automated classifiers are necessary
tools for moderation interventions in platforms attempting to prohibit insults
and textual abuse.  Dai et al.~\cite{olid-task-paper} propose a classifier based
on BERT, a pre-trained language model developed by researchers at
Google~\cite{BERT}.  We include this classifier as it already contained a
language model.  BERT is a medium-size model with hundreds of millions of
parameters, so there is potential for an improvement in performance from
augmenting the model's training data using a larger model.  On the other hand,
the fact that there is already information from a language model being taken
into account may mean that there is less to gain from using another.

The researchers make use of the Offensive Language Identification
Dataset (OLID)~\cite{OLID}, which is a dataset of 14,200 annotated Twitter posts.
There are 9160 (approximately 65\%) negative samples, with the rest being
offensive and sorted into one of four categories based on their target. Every
sample is of English text. Some minor prepossessing has taken place, such as
URLs being shortened to the term ``URL" and references to specific users have
been replaced with the term ``@USER". An open-source implementation of their
model has been made
available\footnote{\href{https://github.com/wenliangdai/multi-task-offensive-language-detection}{https://github.com/wenliangdai/multi-task-offensive-language-detection},
accessed 24/04/2020}.  

We focus on the BERT-based classifier from Dai et al.~\cite{olid-task-paper}
which distinguishes between the two primary classes. This classifier has multiple different models. The most advanced of these uses subclass data to inform its primary class decision.
For simplicity, we used a reduced model with this feature omitted. Even so, when
generating new data we made sure to maintain correct proportions for every
class, not just positive/negative.  

\subsubsection{Task 2: Deceptive Opinion Detection} 
Reviews can be a key tool for consumers to evaluate the quality of products and
services. This makes them a prime target for fraud. Opinion spam, or fake
reviews, can be used to deceive consumers and sway them towards or away from a
purchase.  Salunkhe~\cite{reviews-task-paper} proposes an attention-based
bidirectional LSTM to classify these reviews as either truthful or deceptive.
This model was chosen as it represents a complex deep network topology which
reflects the architecture of many modern classifiers. 

The data used is a balanced set of hotel reviews from a range of sources. The
truthful samples come from multiple review websites, while the negative samples
are from Mechanical
Turk\footnote{\href{https://www.mturk.com/}{https://www.mturk.com/}}. The
samples are described by Ott et al. (2011)~\cite{reviews-data-2} and Ott et
al. (2013)~\cite{reviews-data-1}.  There are a total of 1600 reviews, with 400 for each
combination of opinion polarity and truthfulness.  Salunkhe's open-source
implementation\footnote{\href{https://github.com/ashishsalunkhe/DeepSpamReview-Detection-of-Fake-Reviews-on-Online-Review-Platforms-using-DeepLearning-Architectures}{https://github.com/ashishsalunkhe/DeepSpamReview-Detection-of-Fake-Reviews-on-Online-Review-Platforms-using-DeepLearning-Architectures},
accessed 24/04/2022} contains not only their final classifier but also a
selection of basic ML models, providing an additional dimension of comparison
for this task.
We primarily focus on the advanced classifier which performs best on this task,
but also report results for augmentation under the various `basic' classifiers.
The advanced classifier is a combined CNN LSTM which makes use of both doc2vec
and TF-IDF, as well as multiple preprocessing stages. The classifier itself has
three convolutional layers with dropout and max-pooling between, plus one
bidirectional LSTM layer and a final dense layer.


\subsubsection{Task 3: SMS Spam Detection}

SMS spam messages often attempt to coerce a target into performing some desired
action, such as leading them into an advance fee fraud
scam~\cite{advance-fee-fraud} or opening a link to malware. Manual review of
these messages raises significant privacy concerns, creating a desirable
application area for an automated classifier. Almeida et~al.~\cite{sms-datset}
assembled a dataset of such messages to enable machine-learning classification.
The set contains 5574 messages, 747 (around 13\%) of those being spam. 

Chandra \& Khatri~\cite{sms-task-paper} propose a straightforward LSTM approach for this
task. Their model contains an embedding layer which helps with learning the
relationships between words. This model is similar to that by
Salunkhe~\cite{reviews-task-paper} but does not have any convolutional layers,
making it a pure RNN, rather than a combined CNN-RNN like the opinion spam
classifier.  The model also makes use of pre-trained GloVe~\cite{glove} vectors
for word embeddings. In our reimplementation we use 100-dimensional vectors from
the glove.6B dataset, trained with 6 billion
tokens\footnote{\href{https://nlp.stanford.edu/projects/glove/}{https://nlp.stanford.edu/projects/glove/}}.
This was chosen as it presents an opportunity to compare how small differences
may influence the effectiveness of the technique to be assessed. An open-source
implementation of the model is available
online\footnote{\href{https://github.com/Awesome12-arch/Detecting-the-Spam-messages-using-Keras-in-Python}{https://github.com/Awesome12-arch/Detecting-the-Spam-messages-using-Keras-in-Python},
accessed 24/03/2022}.


\subsection{Generative Model}
\label{sectionLMselection}

We used GPT3~\cite{GPT-3} with the `Curie' model for text generation. This model
takes an arbitrary string as input and attempts to predict what should be written after it. Strings are encoded as lists of tokens, each approximately 4 characters in English.
We experimented with supplying a single token prompt for every completion, similar
to the approach used by Kumar et al.~\cite{genAug2kumar}.  For example, in Task 1 the prompts used for
negative samples were ``NOT'' and untargeted offensive samples were ``UNT''. Qualitative
inspection of samples showed poor performance, and we found that plain natural
language produced better results.The previous examples became
``A regular tweet -$>$" and ``An untargeted offensive tweet -$>$" respectively.
The other classes (targeting an individual, group or other) were updated in the
same way. The addition of a consistent end sequence ``-$>$" reduced failures from the
model trying to write a longer prompt, rather than write a sample after the
prompt.

\subsection{Experimental Procedure}

For each of the three tasks, we begin with a standard dataset and first
replicate previously published state-of-the-art results on the relevant task. 
We then artificially introduce a lack of data into the dataset by removing some
examples from the training set. Similar truncation methods are common for
research in this area~\cite{genAugFirst,GPT-3-aug-recent,genAug3}. 

\begin{figure}[h]
\centerline{\includegraphics[width=.8\textwidth]{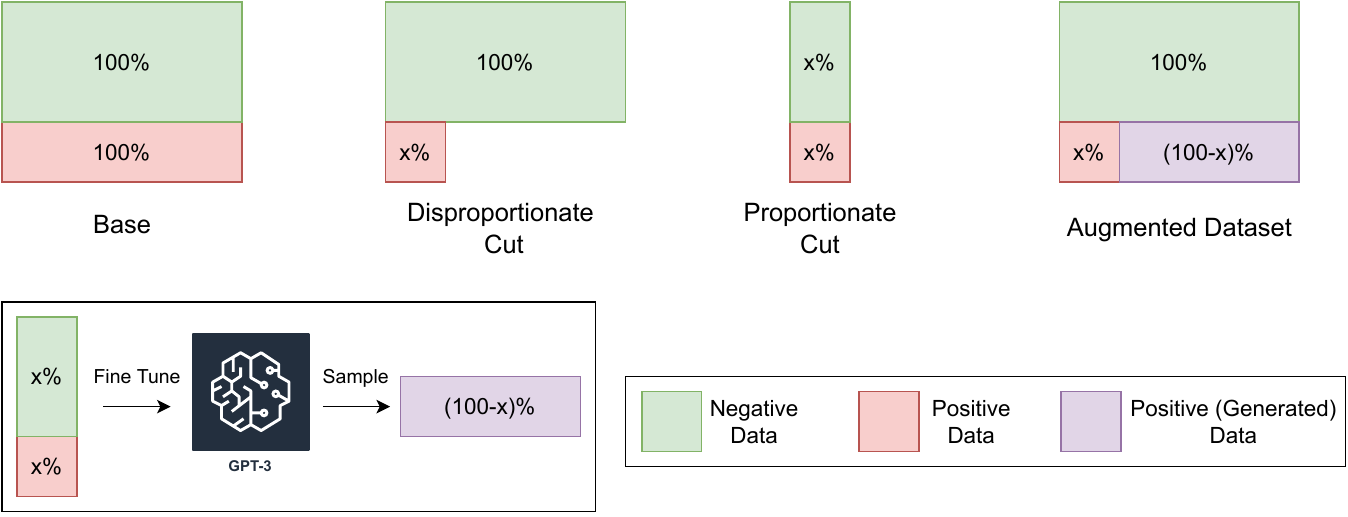}} 
\caption{Truncation and augmentation under proportionate and disproportionate
removal.}
\label{figDatasets1} 
\end{figure}

As illustrated in Figure~\ref{figDatasets1}, two factors are of interest at
this stage. First, the proportion $x$ of the data which is retained after
truncation. We experimented with varying degrees of data removal in each task,
with original data retention at 40, 36, 25, 15, 10, 3 and 1 percent.  The second
factor is whether data is missing proportionately from the positive and negative
classes for each task, or whether data is missing only for the positive class.
In many security contexts, positive labelled data of threats is more difficult
to obtain than negative data.  Accordingly, we explore this
\emph{disproportionate} data limitation, in which data is less available in the
training set than it would normally appear in the dataset (i.e., on top of any
existing class imbalance).  Our training sets are then used as input to a
finetuning process for GPT-3, and samples drawn from this fine-tuned language
model are then used to augment the dataset by replacing the removed data.

\begin{figure}[h]
\centerline{\includegraphics[width=.8\textwidth]{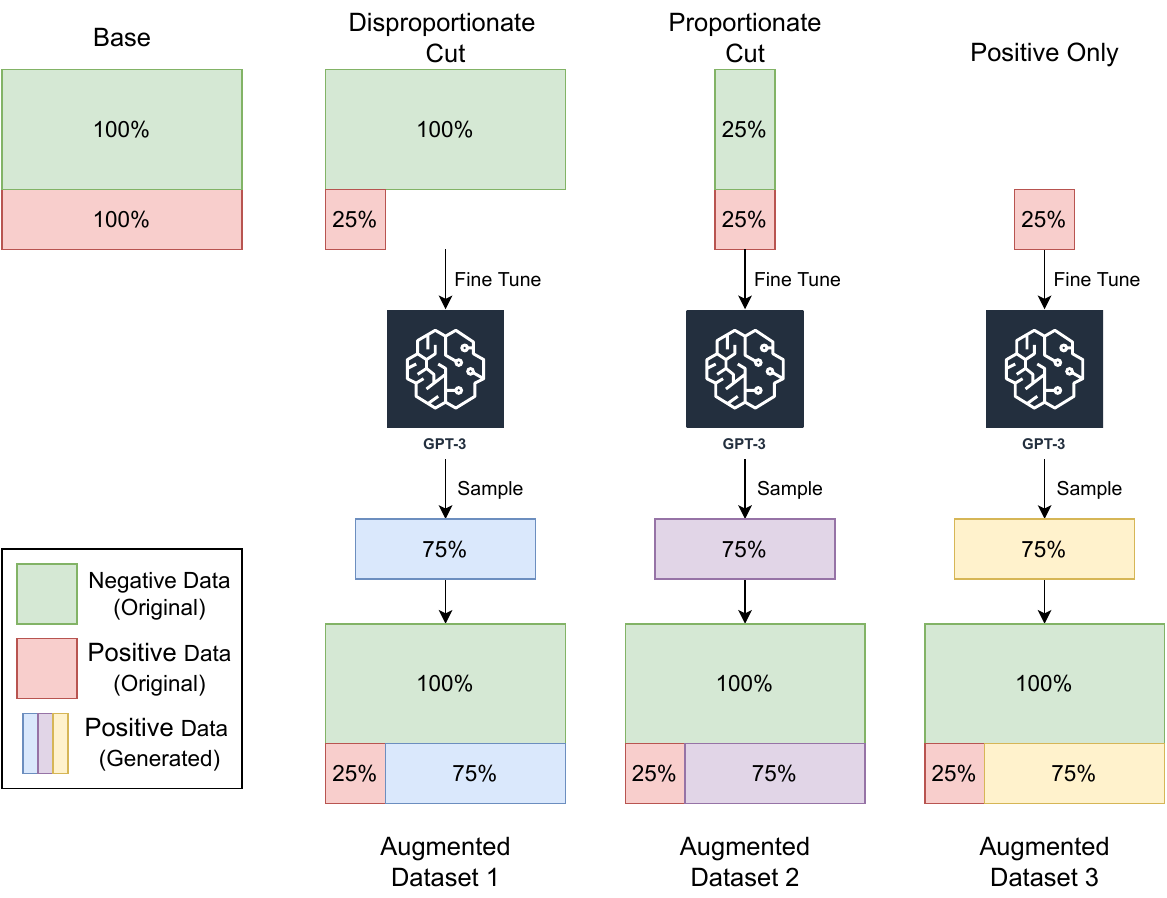}} 
\caption{GPT-3 data augmentation strategies, illustrated with retained data proportion $x =
.25$}
\label{figDatasets2} 
\end{figure}

Figure~\ref{figDatasets2} demonstrates how different fine-tuning approaches are
used to create augmented datasets for a given proportion of missing data. As
well as fine-tuning GPT-3 on the disproportionate and proportionate cases, we
also explore an approach where only positive-class examples are used in
fine-tuning (even though true negative examples are available for forming the
augmented dataset). For any given proportion of retained data $x$, we create and 
evaluate classifiers trained on datasets augmented via these disproportionate, proportionate and
positive-only fine-tuning strategies. 

As points of comparison, we also report the baseline or target performance (of
the full dataset with no data removed), the performance of the classifier
trained on a dataset with data removed but no augmentation, and performance
figures when basic data augmentation techniques (synonym replacement using
WordNet~\cite{wordnet}, word
insertion~\cite{glove} and BERT-guided word insertion~\cite{nlpaug}) are used
instead of GPT-3 text generation to augment the datasets. This use of three
basic augmentation methods as baselines was chosen to match the methodology of
Kumar et al.~\cite{genAug2kumar}.  Table~\ref{tab:shorthand} 
outlines the shorthand used in our results sections.  All model performance is assessed via
the F1 score on the held-out test set (which is never seen by either the
language model or the classifier).

\begin{table}[H]
\centering
\begin{tabular}{ c p{6cm} }
\hline
\texttt{disp} & An unaugmented training set with $x$\% positive, 100\%
negative training examples. \\
\texttt{prop} & An unaugmented training set with $x$\% of positive and negative
samples. \\
\texttt{bda1} & Basic data augmentation using a synonym replacement scheme. \\
\texttt{bda2} & Basic data augmentation using random word insertion. \\
\texttt{bda3} & Basic data augmentation using word insertion guided by BERT. \\
\texttt{gen1} & GPT-3 data augmentation fine-tuned on a disproportionately cut
training set. \\
\texttt{gen2} & GPT-3 data augmentation fine-tuned on a proportionately cut
training set. \\
\texttt{gen3} & GPT-3 data augmentation fine-tuned on only $x$\% positive samples. \\
\hline
\end{tabular}
\caption{Shorthand for the dataset forms used in augmentation experiments, where
$x$ is the proportion of original data retained.}
\label{tab:shorthand}
\end{table}
\clearpage

\section{Results}
\label{sec:results}

\subsection{Task 1: Offensive Language Identification} 


\begin{minipage}[b]{0.55\linewidth}
\centering
\centerline{\includegraphics[width=\textwidth]{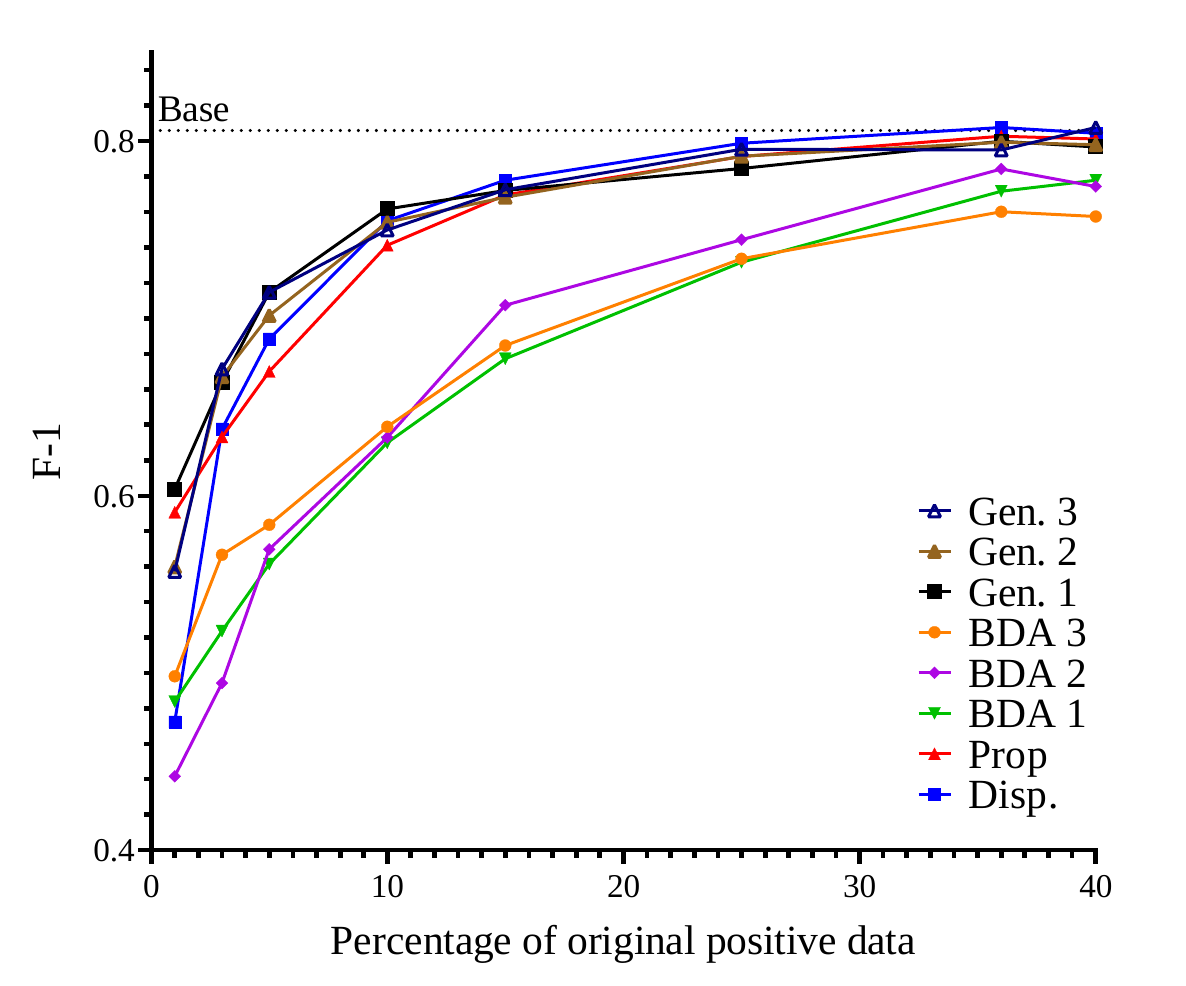}} 
\captionof{figure}{Augmentation strategy performance for offensive language detection in
OLID.}
\label{OLIDresults} 
\end{minipage}\hfill
\begin{minipage}[b]{0.4\linewidth}
    \centering
    \begin{tabular}{|c|c|}
    \hline
     Dataset & Average gap to best \\ \hline
     \texttt{disp} & 3.9\% \\ 
     \texttt{prop} & 2.53\% \\ 
     \texttt{bda1} & 13.76\% \\ 
     \texttt{bda2} & 14.16\% \\ 
     \texttt{bda3} & 12.47\% \\ 
     \texttt{gen1} & 0.74\% \\ 
     \texttt{gen2} & 1.88\% \\ 
     \texttt{gen3} & 1.48\% \\ 
     \hline
    \end{tabular}
    \captionof{table}{The average percentage gap between each dataset and the best performing dataset of each split percentage in Task 1}
    \label{figclasstest-rename}
\end{minipage}

Figure~\ref{OLIDresults} graphs the performance of the different training
strategies over the range of original data retention proportions within OLID. It
is immediately obvious that the basic data augmentations we use as a comparison
are actually harmful to performance in this task, with training sets augmented
using these methods producing \emph{worse} performance than unaugmented training
sets in almost all cases. This may be an example of the poor understanding of
the language by these basic techniques hindering the performance of the BERT-based
classifier by making inappropriate insertions or replacements -- a pitfall
avoided by the GPT3 augmentation strategies.

The overall performance profile shows the impact of data removal on performance,
which is serious at high removal rates.  It is worth noting, however, that at
40\% retention (i.e.  with 60\% removed), even the unaugmented datasets attained
figures close to the 0.806 F1 baseline performance, showing that Dai et al.'s
method is remarkably robust to loss and imbalance. However, performance does
decrease as more data is removed, and, as retention dips below 10\%, the gap
widens between the performance of unaugmented and GPT-3 augmented training sets.
The best performer under the most severe removal condition (1\% of positive
class data remains) is \texttt{gen1}, fine-tuned on a disproportionate training
set. Table~\ref{figclasstest-rename} averages the performance of strategies at each
retention rate, relative to the best-performing strategy at that rate. This
allows for an overall numeric comparison. The disproportionate augmentation of
\texttt{gen1} proves to be the overall best performing strategy, followed by the
other GPT-3 augmentation strategies and then the proportionate unaugmented
dataset strategy.

\subsection{Task 2: Deceptive Opinion Detection}



All mean F1 scores are shown in Figure~\ref{reviewresults}. In short, all
classifiers that were augmented with GPT-3-generated data outperformed all
those that were not. The profile of the GPT-3-augmented strategies
is quite different to the others. This is most clearly seen in
Table~\ref{tabdifference}, which compares the mean F1 scores of the
lowest-performing GPT-3-augmented dataset and highest performing other
dataset, at each split percentage.  

\begin{minipage}[b]{0.55\linewidth}
    \centering
    \begin{adjustbox}{center}
    \includegraphics[width=\textwidth]{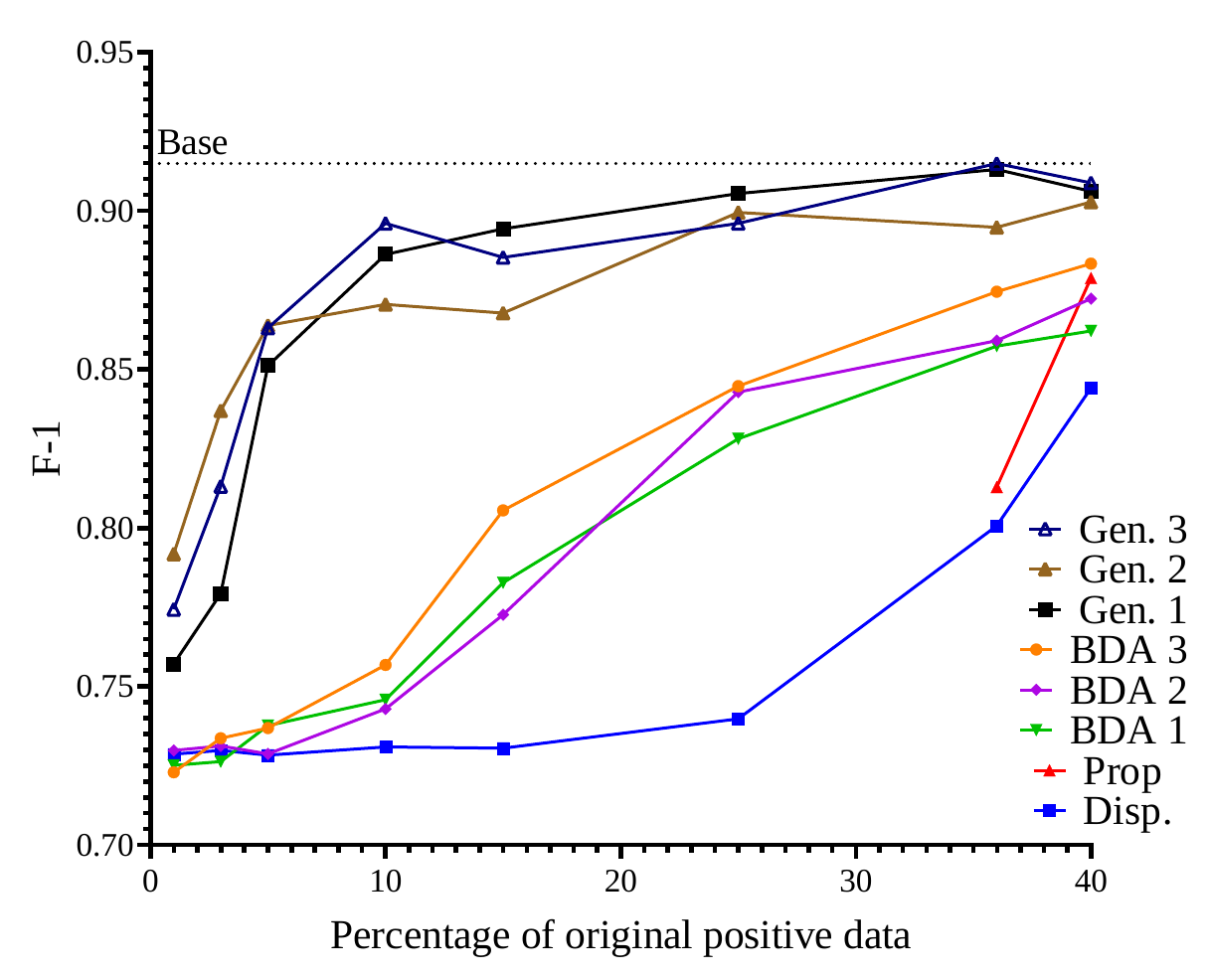}
    \end{adjustbox}
    \captionof{figure}{Augmentation strategy performance for deceptive opinion detection.}
    \label{reviewresults}
\end{minipage}
\begin{minipage}[b]{0.4\linewidth}
    \centering
    \begin{tabular}{|c|c|}
    \hline
     Dataset & Average gap to best \\ \hline
     \texttt{disp} & 13.774\% \\ 
     \texttt{prop} & 7.226\% \\ 
     \texttt{bda1} & 10.567\% \\ 
     \texttt{bda2} & 10.365\% \\ 
     \texttt{bda3} & 9.276\% \\ 
     \texttt{gen1} & 1.654\% \\ 
     \texttt{gen2} & 1.035\% \\ 
     \texttt{gen3} & 1.137\% \\ 
     \hline
    \end{tabular}
    \captionof{table}{The average percentage gap between each dataset and the best
performing dataset of each split percentage in Task 2} \label{figclasstest2}
\end{minipage}

At the lowest of retention of the original positive-class data, the performance
gap is initially minor, at about 0.02, but quickly widens to more than 0.11 at
5\% retention.  The worst generation strategy at this point, \texttt{gen1} is
more than 0.15 F1 improved upon the performance of the unaugmented dataset with
disproportionate losses.  Performance rapidly accelerates with
generator-augmented datasets up to the 10\% retention mark. Up until this point,
the other datasets show almost no improvement.  After this, there is a clear
turning point. The performance of the generator strategies begins to plateau as
positive data retention levels increase, while the other datasets start to
improve.  The gap consistently narrows from then on. 



\begin{table}
\footnotesize
    \centering
    \begin{tabular}{|c|>{\centering\arraybackslash}p{2cm}|>{\centering\arraybackslash}p{2cm}|c|}
    \hline
        Split (\%) & Lowest F1 score of GPT-3 augmented & Highest F1 score of
other approaches & Difference \\\hline
        1   & 0.7571 & 0.7298 & 0.0273\\
        3   & 0.7792 & 0.7336 & 0.0456\\
        5   & 0.8511 & 0.7376 & 0.1135\\
        10  & 0.8704 & 0.7567 & 0.1137\\
        15  & 0.8677 & 0.8055 & 0.0622\\
        25  & 0.8960 & 0.8446 & 0.0514\\
        36  & 0.8947 & 0.8744 & 0.0203\\
        40  & 0.9028 & 0.8833 & 0.0195\\\hline
    \end{tabular}
    \caption{The of F1 scores of different dataset types.}
    \label{tabdifference}
\end{table}

Across the three basic dataset augmentation methods, there were some clear
performance trends. When starting with very few positive samples, namely 1\%,
3\% or, 5\%, the effect on performance was minimal. The maximum difference in
mean F-1 score between the highest and lowest-performing basic augmentation
strategies in the entire range was 0.01 F1.  Once at least 10\% of the positive
samples are retained, a new trend emerges.  All augmentation methods outperform
the disproportionate dataset by a wide margin, until around 40\% positive
samples.  At this point, the augmented sets appear to begin levelling off while
the disproportionate set still trends upwards towards the baseline. Out of the
three basic augmentation strategies, contextual word insertion performs best.
The unguided word insertion and synonym replacement strategies offer comparable
performance. 


Figure~\ref{reviewresults} also shows how performance varies between the GPT-3
fine-tuning strategies used for augmentation. The first two points indicate
performance with only 1\% and 3\% of the original positive data.  Here, the
characteristics are different to the rest of the graph.  There is a clear
hierarchy, with the balanced fine-tune set (\texttt{gen2}) performing best,
followed by the positive-only set (\texttt{gen3}), and then the disproportionate
set (\texttt{gen3}).  A different trend can be observed from 5\% onward. There
is no clear `best' strategy, however the disproportionate and positive-only sets
are generally superior to the balanced set. In the averaged performance figures
in Table~\ref{figclasstest2}, we see that \texttt{gen2} is the overall best
performer when considering different retention rates, but by a much closer
margin than in Task 1.

\subsection{Task 3: SMS Spam}

The average F-1 scores for each dataset at each split percentage are shown in Figure \ref{figlineSMS}. 
The generator-augmented datasets again show the highest performance of all.
Notably, across the 3\%, 5\% and 10\% splits, they hold a clear
advantage over the other strategies. 
This fades as more data becomes available. By 36\%, all datasets become equally viable. 

One interesting aberration in the performance profile is that the top four strategies
lose performance when moving from 3\% to 5\% original positive data.  Given that
they \emph{all} lose performance, the effect may be attributed to the data they
are given. As the data to be cut is randomly selected each time (but constant
across strategies at the same percentage) it may be that the positive data
selected for the 5\% retention cut happened to provide less useful information
than in the 3\% cut. However, the weaker strategies still appeared to benefit
from the increased availability of original positive samples, perhaps because
they suffered more severely from the initial lack of data.

\begin{minipage}[b]{0.55\linewidth}
    \centering
    \begin{adjustbox}{center}
    \includegraphics[width=\textwidth]{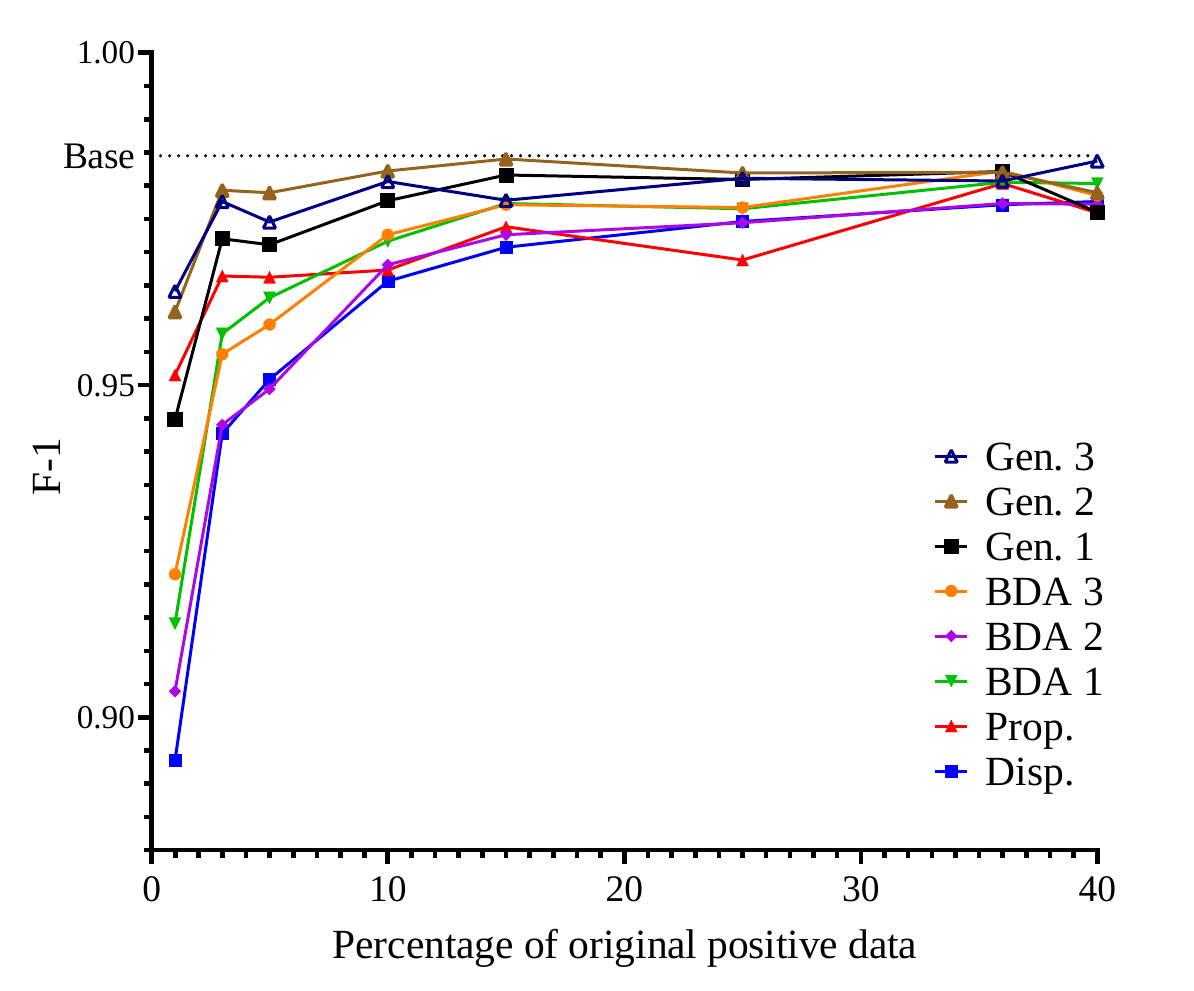}
    \end{adjustbox}
    \captionof{figure}{Augmentation strategy performance for SMS spam detection.}
    \label{figlineSMS}
\end{minipage}
\begin{minipage}[b]{0.4\linewidth}
    \centering
    \begin{tabular}{|c|c|}
    \hline
     Dataset & Average gap to best \\ \hline
     \texttt{disp} & 2.356\% \\ 
     \texttt{prop} & 1.099\% \\ 
     \texttt{bda1} & 1.479\% \\ 
     \texttt{bda2} & 2.171\% \\ 
     \texttt{bda3} & 1.460\% \\ 
     \texttt{gen1} & 0.641\% \\ 
     \texttt{gen2} & 0.102\% \\ 
     \texttt{gen3} & 0.207\% \\ 
     \hline
    \end{tabular}
    \captionof{table}{The average percentage gap between each dataset and the best
performing dataset of each split percentage in task 3} \label{figclasstest3}
\end{minipage}

Table~\ref{figclasstest3} shows that the overall best performing model was
\texttt{gen2}, but all three generator-augmentation strategies are quite tightly
competitive. The disproportionate unaugmented and basic word insertion
strategies performed on par with each other, but worst of all when positive data
was limited to $<$ 10\%. At 10\% they match the effectiveness of the
proportionate dataset, then exceed it at 25\%.  Their performance curve is
similar to that of the other two basic augmentation datasets. The synonym
replacement and contextual insertion datasets follow the same profile but with
improved performance.  The improvement over the disproportionate unaugmented
strategy can be expected, however the distance to the basic word insertion
strategy is more interesting. The gap may reflect the fact that the basic word
insertion augmentation is performed using the same GloVe~\cite{glove} database
as is used in the classifier itself. The model is therefore not gaining any
additional information that it does not already have access to.

 \paragraph{Disproportionality} The unaugmented proportionate strategy shows
unique behaviour; it has much higher variation and a shallower performance curve
than other non-generator-augmented strategies. This contradicts the patterns
seen previously.  An explanation for this may simply be that this model's
performance is more strongly tied to the proportionality of the training data it
receives.  At low percentages, this proportionality supports the model's
performance while others do poorly. As the bias of the disproportionate dataset
decreases, and it is given more positive samples, its performance consistently
trends upwards. 

The strategies using basic dataset augmentation techniques also have
proportionate datasets, however they have similar performance curves to the
disproportionate dataset.  This implies that the basic dataset augmentation
techniques used are hindering performance with low split percentages.
The samples are being duplicated and manipulated too many times for the size of
the dataset. They may therefore see an improvement if basic augmentation
strategies were only used to increase the sample count by a smaller number e.g.
from 3\% to 12\% rather than only to 100\%.

\section{Discussion}
\label{sec:analysis}

\subsection{Generated Data Performance}

Results from all three tasks have shown evidence that modern text generators
using large language models can improve security classifier performance when
used for dataset augmentation.  However, generated data is not a perfect
substitute for true data, and the magnitude of the improvement possible may
depend on the task, the classifier being used, and the quantity of positive
samples available for fine-tuning the language model. Strategies using generator
augmentation would seem to be most helpful when only a small number of true
positive samples are available relative to the expected prevalence in the
testing set or deployment scenario.

The results from Task 1 indicated that provided at least 25\% of the original
positive data was retained, there was negligible improvement from using
generated data.  Tasks 2 and 3 contradicted this, however, finding that
generator-augmented strategies more consistently outperformed unaugmented
strategies across different data retention levels.  It can therefore be inferred
that the structure of either the dataset or classifier made the technique less
effective.  The data in Task 2 is mostly incomparable to that of Task 1, as it
is of a much longer format, and presents varied sentiment in both positive and
negative classes.  The dataset in Task 3 is a closer match, given the short
sample length and conversational nature of the text.  We suspect that the
difference in performance for Task 1 stems from the structure of the classifier,
rather than the dataset.  The classifiers in Tasks 2 and 3 were similar,
containing LSTM components.  The classifier in Task 1 by contrast was based on
the BERT~\cite{BERT} language model. We would suggest that using a language
model within the architecture of the classifier itself produced a degree of
generalisability that reduced the effectiveness of---or need for---dataset
augmentation using another language model.

Across all three tasks, the generator-augmented datasets showed the highest
relative performance when limited to 3-10\% of the original positive data. In
absolute terms, this refers to between 274--916 offensive Tweets (Task 1),
24--80 deceptive reviews (Task 2) and 22--75 unwanted SMS messages (Task 3). Our
experiments thus far do not provide a concrete answer regarding the quantity of
data required for successful application of the technique, but these ranges
could be considered guides for similar tasks where collection of positive
samples may be expensive or difficult.  That there is an optimal range in which
this approach is most effective has an intuitive explanation.  Given too few
samples, the language model will have too little information to sufficiently
ground its generation of class examples.  Conversely, on tasks with bountiful
positive samples and a classifier already leveraging a language model, the
improvements granted by this technique may be small to negligible.

Our results are similar to those found in the work of Kumar et
al.~\cite{genAug2kumar} and Anaby-Tavor et al.~\cite{genAugFirst} in that they
show a clear improvement, particularly in more strongly data-limited scenarios.
They differ however in the magnitude of the improvement. The prior work shows
large gains of up to 40\% in some cases. This is likely reflective of the
baseline performance of the classifiers themselves.  The target F1 score of the
classifier across all three of our tasks rarely dropped below 0.70 (excluding
the basic models in Task 2). For comparison, the unaugmented performance
(measured as mean accuracy) of the classifiers in~\cite{genAug2kumar}
and~\cite{genAugFirst} is usually between 0.4 and 0.6.  They therefore have much
more potential for increase. The tasks chosen in those papers are purposeful
benchmarks, specifically designed to be challenging in order to show differences
between models. By contrast, the classifiers we replicate are representative of
the state of the art in each field.

\subsection{Language Model Fine-Tuning}

We do not see a strongly conclusive result regarding which fine-tuning approach
is most desirable. However, there appeared to be a few trends in how the
fine-tuning strategy used for the language model would influence the
effectiveness of the samples generated.  First, with low quantities of true
data, the best approach appears to be to use a proportionate or positive-only
dataset. It was shown in both Tasks 2 and 3 that fine-tuning with heavily
disproportionate datasets would decrease performance.  This behaviour is not
however ubiquitous.  These methods were also seen to converge when supplied with
more original data.  Second, results from Task 2 appear to indicate that the
proportionate fine-tuning dataset resulted in poorer generated data than the
other generator strategies at higher retention levels (above 5\% retention).
This outcome was not reflected in the results from Tasks 1 and 3.  We are not
certain if this result reflects an inherent structural difference between the
tasks, or merely speaks to the highly-similar performance of all three generator
strategies on Task 2.

\subsection{Practicality}

\subsubsection{Cost Analysis}

 Aside from the cost of running the classifiers themselves, the material costs
associated with this project came from two sources: language model fine-tuning,
and language model sampling (generation).  
Unfortunately, generation without fine-tuning cannot be advised. A small number
of tests were run trialling this method and performance became worse than random
guessing when including the data. 
The price of fine-tuning is dependent on three things: the number of tokens, the
base engine, and the number of training epochs. The price per token is half that
of the base engine cost i.e. \$0.003 per 1000 tokens for Curie, \$0.030 per 1000
tokens for DaVinci. The number of training epochs is up to the user. OpenAI
recommends 4 as standard, so this was used for all instances for this study. 
%
%
All billing was calculated as a function of the action type and the number of
tokens involved in the request.  Text included in the prompt is always included
in the cost calculation at the same rate as generated tokens. 
 The total cost to fine-tune and generate data was relatively low, averaging
approximately \$2 to \$3 for each dataset in this study.  These exact prices and
the associated rules will likely change in the future -- at the beginning of
this project, fine-tuning a model was free, with costs only for generation.

It can generally be concluded that cost should not provide a high barrier for
use of generator-based data augmentation in most instances. The clearest
benefits have been seen when fine-tuning with $<$ 10\% of the original data. At
this point, costs are low and can be less than \$0.10 per dataset.

\subsubsection{Large Language Models as Classifiers}

Task 1 indicated that using a language model as part of a classifier might offer
similar benefits to using this technique. This raises the question: why not just
use a more powerful model like GPT-3 in the classifier? 

At first glance, this suggestion makes some sense. It is quite possible that the
measured increase in performance was at root due to the much larger size and
greater power of the GPT-3 model in comparison to BERT. A similar, or even
larger, improvement in performance may be seen by instead converting the
classifier to make use of GPT-3. The downsides of such an approach stem from
size of the GPT-3 model. With 175 billion parameters, it takes considerable
hardware to even run the model.
Depending on the implementation, BERT has approximately
1000$\times$ fewer parameters. Even so, the Task 1 classifier using it took more
than 10 times longer to train than the others. Considering the
rapidly-increasing scale of new language models, it would be a more efficient use of
resources to purchase a small amount of fine-tuning and generation from an
externally-hosted large language model. The outputs could then be used to augment a training
set for a more lightweight classifier, indirectly passing on some of the
language model's understanding of the dataset.


A second pragmatic reason for taking this approach is that it does not require
the same level of access to the language model. Many large language models are
commercially available through an API, but do not offer source code level
support for being used within a classifier. Easier access to the model's
assistance would be of considerable benefit to researchers and practitioners
working with limited resources.

\subsubsection{Misuse Considerations}

An element of concern for dataset augmentation in security domains---and
especially high-fidelity sample generation as in the models we discuss---is that
any generator of malicious content could also be used as a tool by malicious
actors, in an effort to magnify their impact. Our generator for Task 1, for
example, would be capable of cheaply producing large volumes of abusive
messages, of a nature human annotators would struggle to identify as
automatically generated. More dangerously yet, consider a generator tuned to
augment a dataset of social engineering ploys or mass-market fraud -- compelling
hooks cheaply available at any time to cybercriminals with no skill in the
target domain or even language.

For the moment, this consideration rests with OpenAI, who control access to the
model's API, and monitor accounts for misuse of the service\footnote{We
preemptively explained our own usage of the service to OpenAI to forestall any
such concerns.}.  However, the increasing interest in and availability of
sophisticated text generation capabilities should motivate urgent work to design
defensive classifiers and other solutions capable of protecting internet users
from such risks.

\subsubsection{When to use Generator Augmentation}

As a general guide, augmentation using a text-generating language model will
offer the largest performance improvement when sample counts are extremely
limited, and the classifier itself does not contain a language model. All three
of our evaluation tasks saw the performance gap decrease as true samples were
made available.  This result in disproportionate limitations mirrors the results
obtained by Kumar et al.~\cite{genAug2kumar}, Quteineh et al.~\cite{genAug1},
and Anaby-Tavor et al.~\cite{genAugFirst} in proportionately limited datasets.
It also appears that when very few samples are available, fine-tuning may be
more effective if conducted with a balanced dataset. 

The length of the sample did not appear to have an appreciable impact on
performance. Task 2 had sample lengths up to 4200 characters (750 words) and had
comparable performance to Task 3, in which text samples sometimes had as few as
5 characters apiece. Extremely long samples may cause configuration issues if
they begin to exceed the limit of what is allowed by the language model when
fine-tuning.


\section{Conclusion}
\label{chap:conclusion}

This study has built on the work of Quteineh et al.~\cite{genAug1} and Kumar et
al.~\cite{genAug2kumar} to further examine how text-generator dataset
augmentation can be applied to the security domain.  Tasks were selected to
represent different areas of security classification. For each task, an
open-source classification model from a recently published paper representing
the state-of-the-art was identified and replicated. An array of tests then
explored the value of text-generator dataset augmentation in different
configurations.  We find that for most classifiers, this form of data
augmentation is effective, with classifiers trained with generated data on
average outperforming others across our evaluation scenarios in three tasks.

An overarching objective has been to evaluate this method as a solution to the
common problem of disproportionality of availability in labelled security data.
The effects of different rates of positive-class data limitation have been
explored through a series of experiments across three different classification
tasks. We find that text generation can be especially effective for data
augmentation in cases where positive-class samples are very scarce, a positive
result for domains where collecting such examples may be expensive or difficult.

We also investigate which of three fine-tuning approaches is most effective for
generation. This is an area that has not been explored in other data
augmentation research. We find mixed results that tentatively suggest that using
a proportionate training set for fine-tuning purposes may be more reliable.
Future work will attempt clarify these last results, and further probe the set
of factors which should guide the use of text generation in security
classification tasks.



\section*{Appendices}

\appendix

\section{Other Classifier Results}

The analysis of Task 2 focused on the high-performing primary classification
model of Salunkhe~\cite{reviews-task-paper}. There was, however, a separate set
of basic models included in the original analysis, and we additionally evaluated
each strategy using these classifiers. These models were: K-Nearest Neighbours,
Decision Tree, Multinomial Naive Bayes, Gaussian Naive Bayes, Support Vector
Classification, Logistic Regression, and Stochastic Gradient Descent (SGD). The
SGD result was probabilistic, so the test was repeated 20 times. All others were
deterministic.  These models all use the same train/test splits as the main
classifier, but required an adjustment to the preprocessing pipeline. 

\begin{table}[h]
    \centering
    \begin{tabular}{|c|c|c|}
    \hline
        Strategy & Mean F1 &\begin{tabular}[x]{@{}c@{}}Mean F1\\K-NN removed\end{tabular}  \\\hline
        \texttt{disp}     & 0.7118 & 0.7181 \\
        \texttt{prop}     & 0.6440 & 0.6942 \\
        \texttt{bda1}	  & 0.7114 & 0.7275 \\
        \texttt{bda2}     & 0.7166 & 0.7218 \\
        \texttt{bda3}	  & 0.7190 & 0.7316 \\
        \texttt{gen1}     & 0.6704 & 0.7387 \\
        \texttt{gen2}     & 0.6850 & 0.7563 \\
        \texttt{gen3}     & 0.6847 & 0.7576 \\\hline
    \end{tabular}
    \caption{The mean F1 scores of different augmentation strategies on a range of basic classifiers.}
    \label{tabbasic}
\end{table}

Overall performance on the basic classifiers is shown in Table \ref{tabbasic}.
Taking the mean F1 score of each strategy across each model, the results show
poor scores from the generator-augmented datasets. This is because
the K-nearest neighbour results contradict the rest of the testing
suite.  A different picture emerges with these measurements removed. The three
generator-augmented datasets move from having the 4\textsuperscript{th},
5\textsuperscript{th}, and 6\textsuperscript{th} highest average F-1 to
1\textsuperscript{st}, 2\textsuperscript{nd}, and 3\textsuperscript{rd}.
Within the basic models, it appears that the more complex classifiers give
clearer trends than those with simple structures. Interestingly, these results
also show that the performance of the proportionate unaugmented strategy is,
when averaged across classifiers, inferior to that of the disproportionate
unaugmented strategy.

\section{Proportionate Dataset Performance}

\begin{figure}[h]
    \centering
    \begin{adjustbox}{center}
    \includegraphics[width=\textwidth]{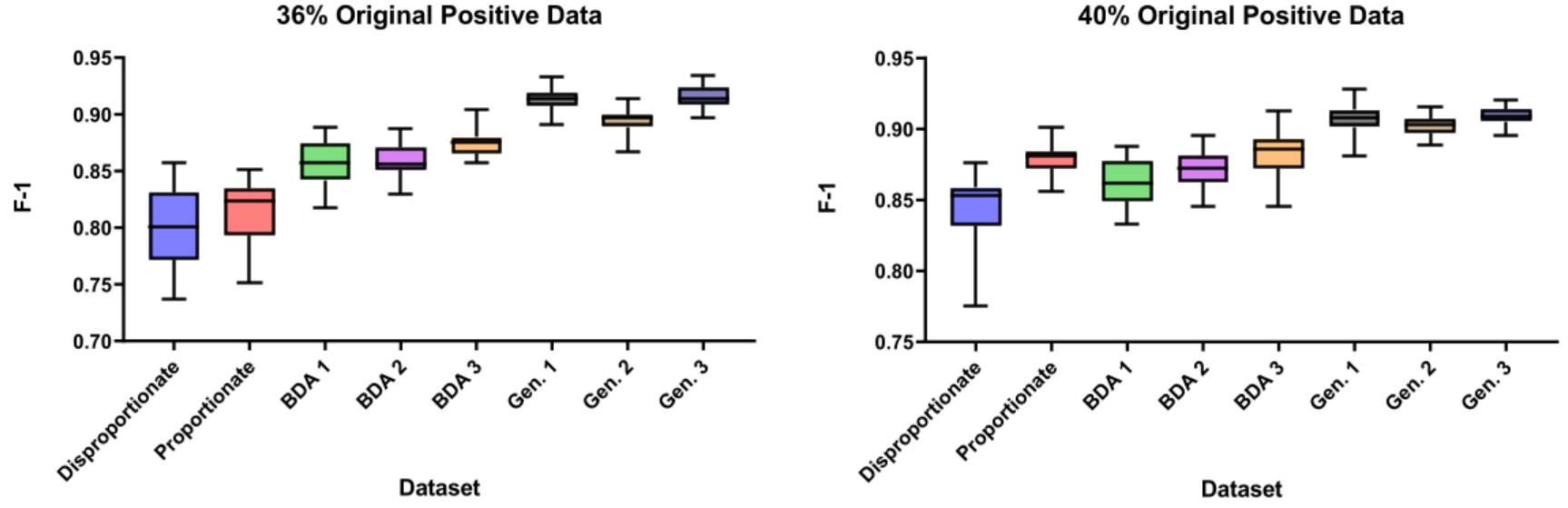}
    \end{adjustbox}
    \caption{F-1 score of of different datasets types with 36\% and 40\% of the original positive data}
    \label{fig3640}
\end{figure}

Unfortunately, for many of the data splits in Task 2 we were unable to use a
proportionately split dataset. This was because the construction of the
classifier required at least 512 samples, and the original small size of the
dataset limited the scenarios for which this was possible to the upper end of
our testing range: 36\% and 40\% of the original training set.  The results for
all tests at these splits are shown in Figure~\ref{fig3640}.  

With 36\% of the original positive data, the proportionate dataset has
performance between that of the disproportionate dataset and those augmented
with basic techniques.  Interestingly, the performance profile is very different
when looking at the 40\% split. The effectiveness of the proportionate dataset
increases substantially, nearly matching that of the best basic augmentation
technique (\texttt{bda3}).  
\end{document}